\documentclass[showpacs,preprintnumbers,amsmath,amssymb,prl]{revtex4}

\usepackage[latin1]{inputenc}
\usepackage{graphicx}
\usepackage{dcolumn}
\usepackage{bm}
\usepackage{color}

\newcommand{\ket}[1]{|#1\rangle}

\newcommand{\0}{|0\rangle}
\newcommand{\1}{|1\rangle}

\begin{document}

\title{Bosonic CNOT gates with Quantum Interrogation.}
\author{Juan Carlos Garc\'ia-Escart\'in}
\email{juagar@tel.uva.es}
\author{Pedro Chamorro-Posada}
\affiliation{Departamento de Teor\'ia de la Se\~{n}al y Comunicaciones e Ingenier\'ia Telem\'atica. Universidad de Valladolid.\\ETSI Telecomunicaci\'on. Camino del Cementerio s/n. 47011 Valladolid, Spain.}
\date{\today}
\begin{abstract}
We put forward a new CNOT gate scheme with atoms and ions based on quantum interrogation and a bosonic particle extension of the models of linear optics quantum computation. We show how the possibility of particle collision can provide the strong interaction that is needed for universal quantum gates. Two atom optics proposals are provided. Unlike previous schemes, these gates are at the same time nondestructive, valid for arbitrary inputs and can work with a probability as close to unity as desired in the lossless case. Data is encoded into position modes and the gates only require basic atom optics elements, which gives potentially simpler quantum computer implementations. 
\end{abstract}

\maketitle
Quantum information processing promises tremendous benefits in the areas of communication and computation \cite{NC00}. There exist many competing proposals for the construction of a working, scalable quantum computer. Two of the most important groups of implementations are those based on the manipulation of individual particles, like neutral atoms or ions, and the schemes with single photons \cite{ARDA04}. 

One surprisingly simple model is the Knill-Laflamme-Milburn, KLM, optical quantum computer based on linear optics \cite{KLM01}. In a recent paper, Popescu has shown that the KLM model is also valid for bosonic particles \cite{Pop07}, combining the best of both the light and matter worlds. While these particles can be equally bosonic neutral atoms or bosonic ions, we will use the term ``atoms'' as a generic shorthand for both. However, unlike in Popescu's proposal, ions can be better suited for some of our schemes. 

By using an adequate arrangement of electric fields and laser beams, atom beams can be manipulated with the same results that are usually obtained for light. Techniques from the growing fields of atom optics and atom interferometry can give atomic equivalents to the optical elements of KLM quantum computation, such as waveguides, beamsplitters and phase shifters \cite{Mey01}. An example are matter wave interferometers, in which light is used as the fixed element that splits, guides and recombines matter waves in a dual of the well-known light interferometers. In the KLM-Popescu model, photons are replaced by flying atoms, either in free propagation or carried inside moving potential wells that follow all the possible branching paths of the wavepacket's trajectory. 

Among the principal advantages of this model is the relative ease with which true single atom sources and efficient atom detection can be implemented when compared to the optical case. This significant advantage apart, the model has the same strengths and weaknesses of KLM computation. In particular, CNOT gates, a key element in any quantum computation implementation proposal, are probabilistic. In a KLM quantum computer, the quantum information units, the qubits, are photons. Individual photons are easy to manipulate and single qubit logic gates can be built, but the interactions between photons that are needed for multiple qubit gates such as the CNOT are extremely challenging to obtain. In KLM computation, the required nonlinear interaction is achieved probabilistically through measurement and there is always a bounded probability of success for a reasonable amount of resources. 

In this paper, we put forward a novel CNOT gate scheme for bosonic KLM-Popescu computation based on quantum interrogation, QI. Our gates offer a suitable alternative to the original KLM and Popescu schemes, which employ the measurement-assisted probabilistic interaction that occurs when multiple photons/atoms impinge on the same beamsplitter. While photonic qubits lack a more efficient interaction mechanism, a strong enough two atom interaction is available in the bosonic particle case in the form of collisions between atoms. However, when two atoms collide, they scatter and unwanted uncontrolled interaction appears. In the KLM-Popescu model, this is avoided by choosing wide enough wave packets, much longer than the atomic scattering length. We will not need to go that far in our proposal. The possibility of a strong scattering interaction is, in fact, useful for a simple QI controlled sign shift, or CZ, which is equivalent to the CNOT gate up to two trivial single qubit gates. With QI, the benefits of this strong interaction can be gained without any actual particle collision.

Quantum interrogation rests on the principles of interaction-free measurement, IFM. IFM allows to extract information from counterfactuals, events that could have happened but, after a measurement, are found not to have occurred. Quantum mechanics allows a formulation that no classical account can give. A striking example is the Elitzur-Vaidman, EV, bomb test in which a highly sensitive bomb that can be triggered by just one photon can be detected without making it explode with an efficiency up to 1/2 \cite{EV93}. Quantum interrogation combines the concepts of IFM and the quantum Zeno effect to obtain almost perfect interaction-free detection with unit probability \cite{KWH95,KWM99}. 

\begin{figure}[ht!]
\centering
\includegraphics{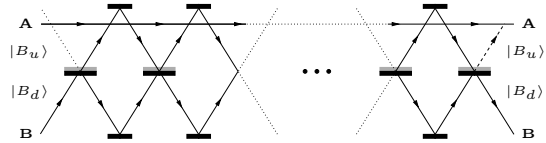}
\caption{Quantum interrogation of a moving bomb in a distributed setup with $N$ beamsplitters.}
\label{bosonic}
\end{figure}

Fig. \ref{bosonic} shows a variation of one of the earliest QI setups \cite{KWH95}, in which the interrogating particle (on port B) was a photon and the bomb was a distributed object so that, if there was a photon in the upper output of any of the beamsplitters, it met the bomb triggering its explosion. We will consider an equivalent case with an electron, instead of a photon, and where the bomb is a moving positron, which will allow later to study QI of quantum objects. In this first scheme, the positron will show no superposition or entanglement and can be treated as a classical bomb.

The setup has $N$ beamsplitters of reflectivity $\cos^2{\theta}$. If we define two states $\ket{B_u}$ and $\ket{B_d}$ for the upper and the lower paths inside the setup, the state evolution at each beamsplitter can be described by an operator $R(\theta)$ such that, 
\begin{center}
$R(\theta)= \left( \begin{array}{rr}
-\cos\theta & \sin\theta \\
\sin\theta & \cos\theta
\end{array} \right)$, with $\ket{B_u}=\left( \begin{array}{r} 1\\0 \end{array} \right)$ and $\ket{B_d}=\left( \begin{array}{r} 0\\1 \end{array} \right).$
\end{center}
The effect of the beamsplitter can be seen as a state rotation by an angle $\theta$. If there is no positron in A, after $N$ beamsplitters, the unperturbed evolution operator is
\begin{center}
$R(\theta)^N= \left( \begin{array}{rr}
-\cos(N\theta) & \sin(N\theta) \\
\sin(N\theta) & \cos(N\theta)
\end{array} \right)$,
\end{center}
as can be easily checked from the spectral decomposition of $R(\theta)$ or by imagining $N$ successive $\theta$ rotations.

If there is a positron in A (bomb case), the evolution is stopped at every stage. We assume that, when the positron and the electron meet, they annihilate and the resulting conversion into energy is detected, in a process equivalent to a measurement in the explosion/no-explosion basis. In this case, the input electron state undergoes the $\ket{B_d}\rightarrow \cos{\theta}\ket{B_d}+ \sin{\theta}\ket{B_u}$ evolution in the first beamsplitter. If there is no explosion, the electron's state is projected back to $\ket{B_d}$. The same evolution is repeated at every beamsplitter and the original input state $\ket{B_d}$ is recovered after each step without explosion. The positron and the electron collide with probability $\sin^2{\theta}$ at each stage. This probability can be made as small as desired by choosing a small enough angle. For small values of $\theta$, there is a probability $\cos^2{\theta}\approx(1-\frac{\theta^2}{2})^2\approx 1-\theta^2$ of avoiding annihilation at each stage. 

For $\theta=\frac{\pi}{2N}$, the electron leaves the setup at the upper port, state $\ket{B_u}$, if there is no bomb, and at the lower port, $\ket{B_d}$, if there is a positron. The global probability of success (interrogation without explosion) after the $N$ stages is $(\cos^{2}\theta)^N\approx 1- N \theta^2=1- \frac{\pi^2}{4N}$, which is negligible for a high enough value of $N$. 

Thus, the state of the electron is altered by the presence or absence of the positron even though their paths cannot have met without triggering their annihilation. Annihilation must be possible for the phenomenon to happen, but the probability of it happening can be made as small as desired, being 0 in the $N\rightarrow\infty$ limit.

When the bomb is also a quantum object, IFM and QI can be used to create entanglement between the interrogating particle and the bomb. The Hardy paradox is one of the first examples of IFM of a quantum object. Hardy's thought experiment examines the consequences of an EV bomb test for an electron and a positron inside two coupled Mach-Zehnder interferometers \cite{Har92}. In the original configuration, the stress was on deriving a new Bell-type experiment to discuss nonlocality in quantum mechanics, but the scheme can also be optimized for entanglement creation using QI. 

QI of quantum objects has been proposed in many schemes as a mechanism for entanglement creation and for the construction of CNOT gates \cite{GWM02}. However, these schemes are either probabilistic, destructive or valid only for a limited range of input qubit states. Qubit encoding will be essential to understand the origin of these limitations. 

We will use, like many other KLM-type schemes, dual-rail encoding. In our path dual-rail encoding, qubits are encoded into a single particle that can follow two alternative paths, one for logical $\ket{0}$ and the other for logical $\ket{1}$. Superpositions of particle paths represent the different qubit values. The CNOT operation on two qubits, for dual-rail, is equivalent to a path switching on the second qubit, called target, whenever the first qubit, called control, is in the path mode that corresponds to $\ket{1}_C$. In the following, subindices C and T will mark control and target qubit states.  

A good example of the problems of the previous proposals is the setup of Fig. \ref{bosonic} as employed by Azuma \cite{Azu03}. The bomb is a now a superposition of positron states and $\theta=\frac{\pi}{2N}$. If $\ket{B_u}\equiv\0_T$ and $\ket{B_d}\equiv\1_T$ and the positron can be outside the setup, $\0_C$, or enter it at port A, $\1_C$, there will be a conditional switching of the paths of the target electron. However, input $\1_C\0_T$ will have such a high probability of explosion that it needs to be forbidden. This makes a limited gate. QI gate proposals offer many ingenious ways of playing down this kind of hindrance. Instead, we will use a more direct route with a new setup with $\theta=\frac{\pi}{N}$. The model is similar to our previous CNOT gates with the QI of quantum particles by light \cite{GC06a}. The bosonic KLM model allows to surmount the difficulties of light to matter coupling at the single atom level that were an obstacle for an easy implementation in those proposals. 

\begin{figure}[ht!]
\centering
\includegraphics{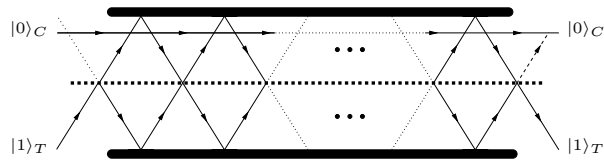}
\caption{Bosonic dual-rail QI CZ gate with two atomic mirrors and a light grating.}
\label{bosonicmirror}
\end{figure}

Our scheme (Fig. \ref{bosonicmirror}) has the same configuration as in Fig. \ref{bosonic}, but electrons and positrons have been substituted for bosonic particles, which provide a scattering interaction equivalent to the annihilation. The detection of scattered atoms is the bomb explosion. The setup can be implemented with only two long atomic mirrors and a grating that is at the same distance from each mirror (a thick discontinuous line in the figure). The mirrors provide the reflection stages and an adequate grating, which can be implemented by stationary light waves, behaves as an atomic beamsplitter for atoms. The target atom only crosses the grating at the points where a beamsplitter is needed.

The bomb will be a moving atom on the upper input. Whenever two atoms cross their paths, they collide and scatter \footnote{The atoms must reach the crossings at the same time. To guarantee that, we can control the speed of the carrier potential well, use media for which the atomic wavepackets have a different speed or choose paths of equal length for both particles.}. The bomb is considered to be the control atom. The first atom, unless scattered, will remain in the same state whether there is a second atom or not. For the no collision case, according to the new $R(\theta)^N$ matrix for $\theta=\frac{\pi}{N}$, the state of the second particle, the target, will be shifted by a $\pi$ phase or not depending on the absence or presence of the first particle. This transformation, which introduces a trivial sign shift for the classical bomb case, becomes a useful way to entangle two particles in the quantum generalization.
 
The setup can be directly employed for a QI CZ gate for bosonic particle qubits. Imagine a detector in each crossing that detects scattering events and two dual-rail bosonic particle qubits such that the $\0_C$ position mode of the control qubit is in the upper path and the $\1_T$ mode of the target qubit follows the lower path.

When there is no scattering, the control qubit is unaffected by the gate. If it is in state $\1_C$, it does not enter the setup. If it is in $\0_C$, it can only be altered on a gate failure, which can be detected. For a $\0_T$ target qubit, the atom does not enter the setup and the qubit state is preserved. For a $\1_T$ target qubit, the particle enters the lower path and acquires a $\pi$ phase shift if it is left to evolve through the $N$ stages. However, a particle on the upper input, if present, acts as a bomb that selects the path through the lower paths with as high as probability as desired. 

If the control qubit is $\1_C$, there is no particle interaction and a sign flip is introduced, but a $\0_C$ control state will prevent the evolution by a continuous projection into the lower subpath of the target qubit. The action on a general input will be the one of a CZ gate, with a sign shift that only happens for the $\1_C\1_T$ input state. The QI CZ gate can be readily converted into a CNOT gate by adding two 50\%, $\theta=\frac{\pi}{4}$, atomic beamsplitters that have at their inputs the two possible paths of the target qubit. These atomic beamsplitters, one before and one after the $N$ splitters that $\1_T$ traverses, act as the two H gates that take a CZ gate into a CNOT. 

The proposed QI CNOT gate is deterministic in the perfect operation regime \footnote{The new total probability of success is $1-\frac{\pi^2}{N^2}$, which tends to 1 in the high $N$ limit.} and, as opposed to Azuma's model, it is valid for arbitrary input qubits because, now, QI only happens on the $\0_C\1_T$ input. Only one mode of the target qubit enters the setup. As a result, it is inadvisable to have more than one output mode, which later would have to be combined with the $\0_T$ path to recover the original dual-rail encoding. This is the reason for the alternative $\theta=\frac{\pi}{N}$, instead of $\theta=\frac{\pi}{2N}$, which produces a controlled sign shift and not a direct CNOT. 

\begin{figure}[ht!]
\centering
\includegraphics{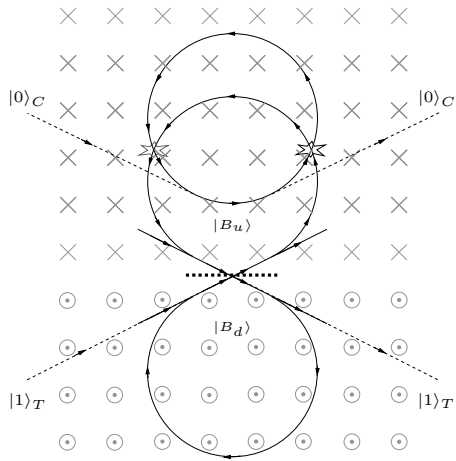}
\caption{Cyclotron setup of a cycling QI CZ gate for positive ions.}
\label{cyclotron}
\end{figure}

We can also give a cycling implementation like that of Kwiat \cite{KWM99}. Ions can be trapped inside circular loops by turning on a constant magnetic field and be later released after $N$ cycles. Fig. \ref{cyclotron} shows a possible cyclotron implementation of QI for positive ions. The atomic beamsplitter (the thick discontinuous line) is supposed to be a light grating. 

The loop configuration gives a simple QI setup for a high number of stages. Ions can be put in and out the loop just by turning a magnetic field on or off, as opposed to the optical case, where switching was a major source of losses.

These gates, like in Popescu's model, are different to most of the other quantum computing proposals with ions. The information is carried in the motional state of the particles and not in their internal levels, which can simplify an experimental realization. The stability of long trajectories, when many gates are present, can be a problem. Nevertheless, the loop configuration should be relatively easy to scale when compared to alternative proposals. 

All the presented bosonic gates are interaction-free inasmuch as the atoms cannot have met under any classical account. In a correct operation, the atoms never collide. Still, they must be given the chance to interact, the more efficiently the better. The scattering efficiency should be ideally 1. Atomic wavepackets of a width similar to the scattering length have high probabilities of colliding. Ions, which have a stronger interaction and a higher scattering cross-section than neutral atoms, are better suited for QI, in contrast to what happened in Popescu's model. If needed, wavepackets can also be broadened and narrowed at different points of the setup using dispersion management schemes analogous to those of photon optics \cite{ETA03}.

Both the cycling and the mirror implementations of the CNOT gate have two crossings in each round. In an ideal operation, ions collide at the first crossing. In some occasions, however, two crossing wavepackets will not be scattered. The additional crossing might be thought of as a second chance. More sophisticated schemes can be designed to increase the number of crossings per round. For instance, the grating at the cyclotron can be turned on and off, by controlling the lasers, so that each QI cycle comprises several rounds inside the loop, with two additional opportunities for scattering per new round. Similarly, QI can work, at a pinch, for smaller absorption rates if the number of cycles $N$ is increased \cite{Jan99}. Nevertheless, there will be a limit to the number of cycles. Longer total trajectories will make the atomic qubits more vulnerable to errors, path instabilities, losses and decoherence. 

Looking at previous realizations of the Hardy though experiment, we can also propose alternative CNOT gates. The internal degrees of freedom of ions have already been suggested for Hardy schemes \cite{Mol01} and the existing optical implementations \cite{LS05} can serve as the basis for the long sought optical CNOT gate. Two-photon absorption, TPA, in a medium which lets single photons pass but absorbs photon pairs, can take the place of the electron-positron annihilation or the ion scattering. TPA is, in fact, central to Zeno optical gate proposals \cite{FJP04}. These gates are equivalent to the CNOT operation inside a modified KLM scheme in which the the failures associated to photon pairs are suppressed using the Zeno effect. However, TPA is a nonlinear optical effect and it will probably be less efficient a mechanism than atomic collision, which produces equivalent effects but is usually stronger.

CNOT gates, when combined with single qubit gates, can provide any possible quantum logic operation \cite{BBC95}. Since single atoms behave exactly like single photons at their corresponding beamsplitters, one qubit gates are also straightforward to build in the atomic case. The proposed CNOT gate is all that is needed for a complete deterministic general bosonic quantum computer. These bosonic quantum computers can have advantages over previous atomic and photonic proposals. The qubits are encoded on the position of the centre of mass of the atomic wavepacket and the gates follow the simple models of linear optics, which should be less experimentally demanding than other atomic implementations that encode qubits in the internal degrees of freedom of the atoms. Sources and detectors are more efficient than in their optical counterparts and collisions give a practical possibility for a CNOT gate. The plethora of experimental atomic techniques makes it difficult to predict a concrete incarnation, but the presented model shows that quantum computation with bosonic atoms is a potential alternative for simple quantum computers. 

This work has been funded by the TEC2007-67429-c02-01 project of the MCyT of Spain. 
\newcommand{\noopsort}[1]{} \newcommand{\printfirst}[2]{#1}
\newcommand{\singleletter}[1]{#1} \newcommand{\switchargs}[2]{#2#1}


\begin{thebibliography}{27}
\expandafter\ifx\csname natexlab\endcsname\relax\def\natexlab#1{#1}\fi
\expandafter\ifx\csname citenamefont\endcsname\relax
  \def\citenamefont#1{#1}\fi
\expandafter\ifx\csname url\endcsname\relax
  \def\url#1{\texttt{#1}}\fi
\expandafter\ifx\csname urlprefix\endcsname\relax\def\urlprefix{URL }\fi
\providecommand{\bibinfo}[2]{#2}
\providecommand{\eprint}[2][]{\url{#2}}
 
\bibitem[{\citenamefont{Nielsen and Chuang}(2000)}]{NC00}
\bibinfo{author}{M.~A. Nielsen} {and}
  \bibinfo{author}{I.~L. Chuang},
  \emph{\bibinfo{title}{Quantum Computation and Quantum Information}},
  \bibinfo{publisher}{Cambridge University Press},
  \bibinfo{address}{Cambridge, UK}, (\bibinfo{year}{2000}).

\bibitem[{\citenamefont{{ARDA experts panel}}(2004)}]{ARDA04}
\bibinfo{author}{{{ARDA experts panel}}}, \bibinfo{journal}{Report
  LA-UR-04-1778, ARDA, available at http://qist.lanl.gov.}
  (\bibinfo{year}{2004}). \bibinfo{note}{Particularly, sections 6.2 (ion
  traps), 6.3 (neutral atoms) and 6.5 (optical)}.

\bibitem[{\citenamefont{Knill et~al.}(2001)\citenamefont{Knill, Laflamme, and
  Milburn}}]{KLM01}
\bibinfo{author}{E. Knill},
  \bibinfo{author}{R. Laflamme} {and}
  \bibinfo{author}{G. Milburn},
  \bibinfo{journal}{Nature} \textbf{\bibinfo{volume}{409}}, \bibinfo{pages}{46}
  (\bibinfo{year}{2001}).

\bibitem[{\citenamefont{Popescu}(2007)}]{Pop07}
\bibinfo{author}{S. Popescu},
  \bibinfo{journal}{Phys. Rev. Lett.} \textbf{\bibinfo{volume}{99}},
  \bibinfo{eid}{130503} (\bibinfo{year}{2007}).

\bibitem[{\citenamefont{Meystre}(2001)}]{Mey01}
\bibinfo{author}{P. Meystre},
  \emph{\bibinfo{title}{Atom optics}}, \bibinfo{publisher}{Springler-Verlag},
  \bibinfo{address}{New York, USA}, (\bibinfo{year}{2001}).

\bibitem[{\citenamefont{Elitzur and Vaidman}(1993)}]{EV93}
\bibinfo{author}{A.~C. Elitzur} {and}
  \bibinfo{author}{L. Vaidman},
  \bibinfo{journal}{\mbox{Foundations of Physics}}
  \textbf{\bibinfo{volume}{23}}, \bibinfo{pages}{987} (\bibinfo{year}{1993}).

\bibitem[{\citenamefont{Kwiat et~al.}(1995)\citenamefont{Kwiat, Weinfurter,
  Herzog, Zeilinger, and Kasevich}}]{KWH95}
\bibinfo{author}{P. Kwiat},
  \bibinfo{author}{H. Weinfurter},
  \bibinfo{author}{T. Herzog},
  \bibinfo{author}{A. Zeilinger}
  {and} \bibinfo{author}{M.~A. Kasevich}, \bibinfo{journal}{Phys. Rev. Lett.}
  \textbf{\bibinfo{volume}{74}}, \bibinfo{pages}{4763} (\bibinfo{year}{1995}).

\bibitem[{\citenamefont{Kwiat et~al.}(1999)\citenamefont{Kwiat, White,
  Mitchell, Nairz, Weihs, Weinfurter, and Zeilinger}}]{KWM99}
\bibinfo{author}{P.~G. Kwiat},
  \bibinfo{author}{A.~G. White},
  \bibinfo{author}{J.~R. Mitchell},
  \bibinfo{author}{O. Nairz},
  \bibinfo{author}{G. Weihs},
  \bibinfo{author}{H. Weinfurter}
  {and} \bibinfo{author}{A. Zeilinger},
  \bibinfo{journal}{Phys. Rev. Lett.}
  \textbf{\bibinfo{volume}{83}}, \bibinfo{pages}{4725} (\bibinfo{year}{1999}).

\bibitem[{\citenamefont{Hardy}(1992)}]{Har92}
\bibinfo{author}{L. Hardy},
  \bibinfo{journal}{Phys. Rev. Lett.}
  \textbf{\bibinfo{volume}{68}}, \bibinfo{pages}{2981} (\bibinfo{year}{1992}).

\bibitem[{\citenamefont{Gilchrist et~al.}(2002)\citenamefont{Gilchrist, White,
  and Munro}}]{GWM02}
\bibinfo{author}{A. Gilchrist},
  \bibinfo{author}{A.~G. White} {and}
  \bibinfo{author}{W.~J. Munro},
  \bibinfo{journal}{Phys. Rev. A} \textbf{\bibinfo{volume}{66}},
  \bibinfo{pages}{012106} (\bibinfo{year}{2002}).
\bibinfo{author}{H. Azuma},
  \bibinfo{journal}{Phys. Rev. A} \textbf{\bibinfo{volume}{70}},
  \bibinfo{eid}{012318} (\bibinfo{year}{2004}).
\bibinfo{author}{M. Pavi\v{c}i\'c},
  \bibinfo{journal}{Phys. Rev. A} \textbf{\bibinfo{volume}{75}},
  \bibinfo{eid}{032342} (\bibinfo{year}{2007}).


\bibitem[{\citenamefont{Azuma}(2003)}]{Azu03}
\bibinfo{author}{H. Azuma},
  \bibinfo{journal}{Phys. Rev. A} \textbf{\bibinfo{volume}{68}},
  \bibinfo{pages}{022320} (\bibinfo{year}{2003}).

\bibitem[{\citenamefont{Garc\'ia-Escart\'in and Chamorro-Posada}(2006)}]{GC06a}
\bibinfo{author}{J.~C. Garc\'ia-Escart\'in}
  {and}
  \bibinfo{author}{P. Chamorro-Posada},
  \bibinfo{journal}{Quantum Inf. Comput.}
  \textbf{\bibinfo{volume}{6}}, \bibinfo{pages}{495} (\bibinfo{year}{2006}).
\bibinfo{author}{J.~C. Garc\'ia-Escart\'in},
  \bibinfo{journal}{Ph.D. thesis, Universidad de Valladolid}, (\bibinfo{year}{2008}).

\bibitem[{\citenamefont{Eiermann et~al.}(2003)\citenamefont{Eiermann,
  Treutlein, Anker, Albiez, Taglieber, Marzlin, and Oberthaler}}]{ETA03}
\bibinfo{author}{B. Eiermann},
  \bibinfo{author}{P. Treutlein},
  \bibinfo{author}{T. Anker},
  \bibinfo{author}{M. Albiez},
  \bibinfo{author}{M. Taglieber},
  \bibinfo{author}{K.-P. Marzlin}
  {and} \bibinfo{author}{M.~K. Oberthaler}, \bibinfo{journal}{Phys. Rev. Lett.}
  \textbf{\bibinfo{volume}{91}}, \bibinfo{pages}{060402}
  (\bibinfo{year}{2003}).

\bibitem[{\citenamefont{Jang}(1999)}]{Jan99}
\bibinfo{author}{J.-S. Jang},
  \bibinfo{journal}{\mbox{Phys. Rev. A}} \textbf{\bibinfo{volume}{59}},
  \bibinfo{pages}{2322} (\bibinfo{year}{1999}).
\bibinfo{author}{G. Krenn},
  \bibinfo{author}{J. Summhammer},
  {and} \bibinfo{author}{K. Svozil},
  \bibinfo{journal}{\mbox{Phys. Rev. A}} \textbf{\bibinfo{volume}{61}},
  \bibinfo{pages}{052102} (\bibinfo{year}{2000}).
\bibinfo{author}{G. Mitchison} {and}
  \bibinfo{author}{S. Massar},
  \bibinfo{journal}{\mbox{Phys. Rev. A}} \textbf{\bibinfo{volume}{63}},
  \bibinfo{pages}{032105} (\bibinfo{year}{2001}).
\bibinfo{author}{H. Azuma},
  \bibinfo{journal}{Phys. Rev. A} \textbf{\bibinfo{volume}{74}},
  \bibinfo{pages}{054301} (\bibinfo{year}{2006}).

\bibitem[{\citenamefont{M{\o}lmer}(2001)}]{Mol01}
\bibinfo{author}{K. M{\o}lmer},
  \bibinfo{journal}{Phys. Lett. A} \textbf{\bibinfo{volume}{292}},
  \bibinfo{pages}{151} (\bibinfo{year}{2001}).

\bibitem[{\citenamefont{Lundeen et~al.}(2005)\citenamefont{Lundeen, and
  Steinberg}}]{LS05}
\bibinfo{author}{J. Lundeen} {and}
  \bibinfo{author}{A. Steinberg},
  \bibinfo{journal}{QELS'05 Conference}, p.~\bibinfo{pages}{310-312} (\bibinfo{year}{2005}).

\bibitem[{\citenamefont{Franson et~al.}(2004)\citenamefont{Franson, Jacobs, and
  Pittman}}]{FJP04}
\bibinfo{author}{J.~D. Franson},
  \bibinfo{author}{B.~C. Jacobs},
  {and} \bibinfo{author}{T.~B. Pittman}, \bibinfo{journal}{Phys. Rev. A}
  \textbf{\bibinfo{volume}{70}}, \bibinfo{eid}{062302} (\bibinfo{year}{2004}).
\bibinfo{author}{P.~M. Leung} {and}
  \bibinfo{author}{T.~C. Ralph},
  \bibinfo{journal}{Phys. Rev. A}
  \textbf{\bibinfo{volume}{74}}, \bibinfo{eid}{062325} (\bibinfo{year}{2006}).
\bibinfo{author}{J. Franson},
  \bibinfo{author}{T. Pittman} {and}
  \bibinfo{author}{B. Jacobs},
  \bibinfo{journal}{JOSA B}
  \textbf{\bibinfo{volume}{24}}, \bibinfo{pages}{209} (\bibinfo{year}{2007}).

\bibitem[{\citenamefont{Barenco et~al.}(1995)\citenamefont{Barenco, Bennett,
  Cleve, DiVincenzo, Margolus, Shor, Sleator, Smolin, and Weinfurter}}]{BBC95}
\bibinfo{author}{A. Barenco},
  \bibinfo{author}{C.~H. Bennett},
  \bibinfo{author}{R. Cleve},
  \bibinfo{author}{D.~P. DiVincenzo},
  \bibinfo{author}{N. Margolus},
  \bibinfo{author}{P. Shor},
  \bibinfo{author}{T. Sleator},
  \bibinfo{author}{J.~A. Smolin} {and}
  \bibinfo{author}{H. Weinfurter},
  \bibinfo{journal}{Phys. Rev. A} \textbf{\bibinfo{volume}{52}},
  \bibinfo{pages}{3457} (\bibinfo{year}{1995}).

\end{thebibliography}
\end{document}